\documentclass[a4paper,11pt]{article}
\usepackage{jheppub} 
\usepackage{lineno}
\usepackage{graphicx}
\usepackage{dcolumn}
\usepackage{bm}
\usepackage{amssymb}
\usepackage{amsmath}
\usepackage{epsfig}    
\usepackage{cancel}
\usepackage{color}
\usepackage{multirow}
\usepackage{slashed}
\usepackage{hhline}
\usepackage{subfigure}

\usepackage{hyperref}
\hypersetup{
    breaklinks=true
}

\def\be{\begin{equation}}
\def\ee{\end{equation}}
\newcommand{\bea}{\begin{eqnarray}}
\newcommand{\eea}{\end{eqnarray}}
\newcommand{\nn}{\nonumber}

\newcommand{\tabincell}[2]{\begin{tabular}{@{}#1@{}}#2\end{tabular}}

\usepackage[T1]{fontenc}
\usepackage{slashed}
\usepackage[utf8]{inputenc}
\usepackage{xspace}
\usepackage{ulem}
\usepackage{array}
\usepackage{ulem,fancyvrb}
\usepackage{xcolor}
\usepackage{makecell}
\usepackage{booktabs}

\newcolumntype{C}{@{\hspace{2em}}c@{\hspace{2em}}}

\newcommand{\tb}{t_\beta}

\newcommand{\cba}{c_{\beta-\alpha}}
\newcommand{\sba}{s_{\beta-\alpha}}

\title{\boldmath Probing the electroweak $4b + \ell + {\rlap{\,/}{E}_T}$ final state in type I 2HDM at the LHC}

\author{Prasenjit Sanyal and Daohan Wang}
\affiliation{Department of Physics, Konkuk University, Seoul 05029, Republic of Korea}
\emailAdd{prasenjit.sanyal01@gmail.com}
\emailAdd{wdh9508@gmail.com}

\abstract{
 Most of the experimental searches of the non-Standard Model Higgs boson(s) at the LHC rely on the QCD induced production modes. However, in some beyond Standard Model frameworks, the additional Higgs bosons can have fermiophobic behaviour. The type I two Higgs doublet model considered here is a perfect example where all the additional Higgs bosons exhibit fermiophobic nature over a large region of parameter space. Thus the electroweak productions of these new Higgs bosons are more dominant over the QCD induced processes. In scenarios with light pseuodoscalar ($A$) which is bound to decay dominantly to $b\bar{b}$, even being fermiophobic, the $4b + W$ state via $p p \to H^\pm A \to (AW)A \to 4b  +  W$ and followed by the leptonic decay of $W$ boson can surpass the QCD initiated $4b$ final state. However, the signal gets overshadowed by large $t\bar{t}+$jets background and hence constructing a suitable discriminator based on the signal hypothesis and signal topology is necessary. We devised a $\chi^2$ variable as the most suitable signal-background discrimintor to reduce the background by a sizable amount and showed the discovery reach ( $>3\sigma$) of the electroweak initiated $4b+ \ell + {\rlap{\,/}{E}_T}$ final state at the LHC.  
}

\begin{document}
\maketitle
\flushbottom

\section{Introduction}
\label{sec:introduction}

 The Standard Model (SM) is currently the most extensively tested model for particle physics as all the collider experiments are in perfect agreement with it. The discovery of Higgs boson \cite{ATLAS:2012yve, CMS:2012qbp, ATLAS:2013dos, CMS:2013btf} gives SM the stature of the only acceptable theory in the energy scale of the present day colliders. However, despite its success, SM is believed neither a complete, nor a perfect theory because of several drawbacks. SM cannot explain the matter-antimatter asymmetry, neutrino oscillation, existence of dark matter, mass hierarchy within elementary particles and does not include gravity. Also there is no fundamental reason in support of one scalar particle, i.e. the Higgs sector in the SM is assumed to be minimal under the electroweak gauge symmetry. The observation of any other scalar particle, whether neutral or charged, will provide a strong indication of the non-minimal framework for the Electroweak Symmetry Breaking (EWSB).
 
  The two Higgs doublet model (2HDM) \cite{Branco:2011iw, Gunion:2002zf, Gunion:1989we, Davidson:2005cw} is the simplest extension of SM which accomodates a second Higgs doublet with the same quantum numbers as the first one. The most general 2HDM where both the Higgs doublets couple to the fermions suffer from Flavor Changing Neutral Currents (FCNC) which contradicts with the experiments. To prevent the FCNC, a $\mathbb{Z}_2$ symmetry is imposed \cite{Glashow:1976nt, Paschos:1976ay} which restricts the Yukawa sector in to four possible types, viz type I, type II, type X and type Y. Each type of 2HDM has its own charecteristics, which makes 2HDM a phenomenologically very rich extension of SM. After the EWSB, the scalar sector in the $CP$ conserving framework of 2HDM  consists of two $CP$ even scalars $h,~H$, a $CP$ odd scalar (pseudoscalar) $A$ and a pair of charged Higgs $H^\pm$. We follow the standard mass hierarchy where the observed 125 GeV Higgs boson, identified by $h$, is the lightest $CP$ even Higgs boson. 
 
The conventional search strategies to look for neutral single Higgs or multi-Higgs states at the Large Hadron Collider (LHC) involves the QCD induced processes like gluon fusion process or the $b\bar{b}$-annihilation \cite{Arhrib:2008pw, Hespel:2014sla, Dicus:1998hs, Balazs:1998sb, Harlander:2003ai, Duhr:2020kzd,CMS:2018qmt, ATLAS:2015rsn, CMS:2019mij, ATLAS:2019odt, CMS:2018hir, ATLAS:2020zms, CMS:2019hvr}, where the $b$-quarks are themselves produced from a (double) gluon splitting. Even for single production of the charged Higgs, the usual search for light charged Higgs is given by the top quark pair production cross section times the branching ratio of top into charged Higgs and the top-bottom asscociated production for heavy charged Higgs \cite{Flechl:2014wfa, Degrande:2015vpa, CMS:2019bfg, ATLAS:2021upq}. Thus all these channels are intrinsically $gg$-induced.  However, the QCD induced processes can be subdominant in some beyond SM (BSM) frameworks where the fermionic couplings to the BSM particles are suppressed. The type I 2HDM fits this criteria as all the BSM Higgs bosons show fermiophobic behaviour \cite{Akeroyd:1995hg, Akeroyd:1998ui, Arhrib:2016wpw, Enberg:2018pye, Kling:2020hmi, Wang:2021pxc, Bahl:2021str, Arhrib:2021xmc, Mondal:2021bxa, Kim:2022nmm, Kim:2023lxc}. In such situation the production of the BSM Higgses through EW processes get more importance than the QCD induced processes \cite{Enberg:2018pye, Arhrib:2021xmc, Mondal:2021bxa}. Similarly, the bosonic decays of the BSM Higgses can overcome the fermionic decay modes if kinematically allowed \cite{Arhrib:2016wpw, Kling:2020hmi, Mondal:2021bxa}. In type I 2HDM, the EW production of a light $A$ in association with $H^\pm$ and followed by $H^\pm$ decay to $AW$ can give the $AAW$ mode. Since the $AAW$ mode involves $q\bar{q}'$-induced production where $q$ represents valence quark ($u,d$), the $AAW$ mode has no QCD production counterpart. For the situation of light $A$, the decays of $A$ are restricted only to the fermions and the branching ratio of $A \to b\bar{b}$ is dominant\footnote{We also have $A\to gg$ decay mode via top/bottom loop, but very suppressed}. This gives rise to the $4b+W$ mode. Considering the leptonic decay of the $W$ boson, we can have $4b + \ell + {\rlap{\,/}{E}_T}$ final state, which as mentioned before is the signature state characteristic of the EW processes and can dominate over the QCD induced $4b$ state for a large parameter space. The $4b + \ell + \rlap{\,/}{E}_T$ final state also have subdominant contributions from other EW processes e.g. through the production channels $p p \to H^\pm H/ H^+ H^- $ and thereafter the decay chains $H^\pm \to AW$ and $H \to AA/AZ$. Contribution to $4b+W$ through $pp \to H^\pm h$ with $H^\pm \to hW$ and $h\to b\bar{b}$ are highly suppressed due to the alignment limit and hence not considered in our work. Since we are restricted to the standard mass hierarchy with light $A$, only $A\to b\bar{b}$ decay can give large cross section for the final state of our interest. Situation with inverted mass hierarchy with $125$ GeV Higgs as the heavier $CP$ even Higgs, identified by $H$, the light $CP$ even Higgs $h$ can decay dominantly to $b\bar{b}$ and we can obtain the $4b + W$ state via the $hhW$ production mode \cite{Arhrib:2021xmc, Kang:2022mdy, Li:2023btx}. Interestingly, in the inverted scenario, fermiophobic limit of $h$ gives large $h\to\gamma\gamma/WW^*$ decay modes \cite{Arhrib:2017wmo, Wang:2021pxc, Kim:2023lxc} compared to the $b\bar{b}$ mode. Whereas, slight deviation from the fermiophobic limit of $h$ makes the decay into $b\bar{b}$ and $\gamma\gamma$ modes comparable, leading to $2b2\gamma+W$ final state \cite{Bhatia:2022ugu} which serves as a complementary channel to the $4b+W$ state.         

 At the LHC the dominant SM background for the proposed final state, the $t\bar{t}+$jets background, is significantly higher than the EW $4b + \ell + {\rlap{\,/}{E}_T}$ final state. Hence we require strong selection cuts which can kill the background. In this work we construct a $\chi^2$ variable, which is based on the signal topology and the signal hypothesis (masses of $A$ and $H^\pm$). The $\chi^2$ can be used to discriminate the signal and the background and therefore highly effective in reducing the background significantly without affecting much the signal. The most appropriate use for $\chi^2$ would be for the mass reconstruction of the BSM Higgs bosons \cite{CMS:2017ixp, Wang:2021pxc}, however we can only reconstruct the masses of $A$ and $H^\pm$ via $4b + \ell + {\rlap{\,/}{E}_T}$ final state. To probe the full Higgs spectrum, the EW induced inclusive $4b +X$ final state (where $X$ implies any additional jets even $b$-jets and /or leptons) is more suitable. In our recent study \cite{Mondal:2023wib}, we showed that the EW initiated $4b + X$ can provide simultaneous reconstruction of all the BSM Higgs boson masses. Hence, instead of using $\chi^2$ for the reconstruction of the masses of $A$ and $H^\pm$, we use $\chi^2$ as a selection criteria to discriminate the signal from the background. Along with the $\chi^2$ we use other selection cuts like the asymmetry cut and di-jet $\eta$ separation cuts to obtain the discovery significance of the signal at the 13 TeV LHC with 3000 fb$^{-1}$ luminosity.         

 The article is organized as follows. In Sec.[\ref{Type I 2HDM REVIEW}] we give an overview of type I 2HDM and discuss the fermionic and gauge couplings of the additional Higgs bosons to show the fermiophobic natures as well as the Higgs trilinear couplings. In Sec.[\ref{4b + W: to probe light $A$}], we discuss the signal topology of the $AAW$ mode, it's production cross section at the LHC and the dominant $t\bar{t}+$jets background. We show the theoretical and experimental constraints on the model parameters, the formalism of $\chi^2$ method and the signal-background analysis. Finally we conclude in Sec.[\ref{sec:conclusion}].  

\section{Type I 2HDM review }
\label{Type I 2HDM REVIEW}
The most general scalar potential for the 2HDM is 
\begin{eqnarray}
\mathcal{V}(\Phi_1,\Phi_2) &=& m_{11}^2 \Phi_1^\dagger \Phi_1 + m_{22}^2 \Phi_2^\dagger \Phi_2 - [m_{12}^2 \Phi_1^\dagger \Phi_2 +h.c.]   
+ \frac{1}{2}\lambda_1(\Phi_1^\dagger \Phi_1)^2 \nonumber \\ 
&+& \frac{1}{2}\lambda_2(\Phi_2^\dagger \Phi_2)^2 
+ \lambda_3 (\Phi_1^\dagger \Phi_1)(\Phi_2^\dagger \Phi_2)  
+ \lambda_4 (\Phi_1^\dagger \Phi_2)(\Phi_2^\dagger \Phi_1) \nonumber \\
&+& \Big \lbrace \frac{\lambda_5}{2}(\Phi_1^\dagger \Phi_2)^2 +[\lambda_6(\Phi_1^\dagger \Phi_1) 
+\lambda_7(\Phi_2^\dagger \Phi_2)](\Phi_1^\dagger\Phi_2) +h.c.\Big\rbrace
\label{2HDM potential}
\end{eqnarray}
where $\Phi_{1,2}$ are the two Higgs doublets with hypercharge $Y=1/2$ and the parameters $m_{11}^2,~m^2_{22}$ and $\lambda_{1,2,3,4}$ are real for the scalar potential to be real. The other parameters $m_{12}^2$ and $\lambda_{5,6,7}$ in general can be complex. To avoid the tree level FCNC, a $\mathbb{Z}_2$ symmetry is imposed under which $\Phi_1 \to \Phi_1$ and $\Phi_2 \to -\Phi_2$  which implies $\lambda_{6,7}=0$. However, the $\mathbb{Z}_2$ symmetry is softly broken by the dimensionful parameter $m_{12}^2 \neq 0$. Assuming $CP$ invariant framework, $m_{12}^2$ and $\lambda_5$ are considered real. The two Higgs doublets are parameterized as  
\begin{eqnarray}
\Phi_i = \begin{pmatrix}
H_i^+ \\
\frac{v_i + h_i + i A_i}{\sqrt{2}}
\end{pmatrix}, \quad i=1,2.
\end{eqnarray}
After EWSB, the scalar spectrum consists of two $CP$ even scalars $h$ and $H$, one $CP$ odd pseudoscalar $A$ and a pair of charged Higgs $H^\pm$. The physical mass eigenstates are related to the gauge eigenstates by the following equations
\begin{eqnarray}
\begin{pmatrix}
H \\ h
\end{pmatrix} = \begin{pmatrix}
c_\alpha && s_\alpha\\
-s_\alpha && c_\alpha
\end{pmatrix}
\begin{pmatrix}
h_1 \\ h_2
\end{pmatrix},
\quad
A = -s_\beta A_1 + c_\beta A_2, \quad 
H^\pm = -s_\beta H_1^\pm + c_\beta H_2^\pm
\end{eqnarray}
where $v_{1,2}$ are the vacuum expectation values (VEVs) of the two Higgs doublets such that $v=\sqrt{v_1^2 + v_2^2} = 246$ GeV and we define the parameter $\tb = v_2/v_1$. We use the abbreviations,  $s_\alpha(c_\alpha) = \sin\alpha(\cos\alpha)$, $s_\beta(c_\beta) = \sin\beta(\cos\beta)$, $t_\beta= \tan\beta$, etc. We identify the physical state $h$ as the SM-like observed Higgs boson of mass $125$ GeV as all the Higgs signal strength measurements  \cite{CMS:2020xwi, Buchbinder:2020ovf, ATLAS:2020bhl,CMS:2020zge,ATLAS:2021nsx,CMS:2021gxc,ATLAS:2020syy,ATLAS:2021upe,CMS:2021ugl,ATLAS:2020wny,ATLAS:2020rej,ATLAS:2020fzp,ATLAS:2022ers} are consistent with SM. Throughout the paper we collectively call $H,~A$ and $H^\pm$ as the beyond SM (BSM) Higgs bosons. 

In type I 2HDM the fermion fields transform odd under the $\mathbb{Z}_2$ symmetry and therefore couple only to the second Higgs doublet $\Phi_2$. Hence the quarks and charged leptons get their masses from the VEV of $\Phi_2$. The Yukawa Lagrangian in type I 2HDM can be written as 
\begin{eqnarray}
\mathcal{L}_Y^I = -\bar{Q}_L Y_u \tilde{\Phi}_2 u_R - \bar{Q}_L Y_u \bar{\Phi}_2 d_R - \bar{L}_LY_\ell \Phi_2 \ell_R + h.c.
\end{eqnarray}   
where $\tilde{\Phi}_2 = i\tau_{2}\Phi_2^*$. After the EWSB, the Yukawa Lagrangian in terms of the mass eigenstates is 
\begin{eqnarray}
\mathcal{L}_Y^I &=&
-\sum_{f=u,d,\ell} \frac{m_f}{v}\left(\xi_h^f\bar{f}fh +
\xi_H^f\bar{f}fH - i\xi_A^f\bar{f}f\gamma_5A \right) \nonumber \\ 
&-&\Big\lbrace \frac{\sqrt{2}V_{ud}}{v}\bar{u} \Big(\xi^u_A m_u P_L + \xi^d_A m_d P_R\Big)dH^+ 
+\frac{\sqrt{2}m_\ell}{v} \xi^\ell_A\bar{\nu}_L \ell_R H^+ + h.c. \Big\rbrace
\end{eqnarray}
where $V$ is the CKM matrix and $P_{L,R} = \frac{1}{2}(1\mp\gamma_5)$ are the chirality projection operators. The Yukawa coupling modifiers $\xi^f$ are given in Table.~[\ref{Tab:YukawaFactors}].
\begin{table}[t]
\setlength{\tabcolsep}{2pt}
\begin{center}
\begin{tabular}{|c||c|c|c|c|c|c|c|c|c|}
\hline
~2HDM~& ~$\xi_h^u$~ & ~$\xi_h^d$~ & ~$\xi_h^\ell$~
& ~$\xi_H^u$~ & ~$\xi_H^d$~ & ~$\xi_H^\ell$~
& ~$\xi_A^u$~ & ~$\xi_A^d$~ & ~$\xi_A^\ell$~ \\ \cline{2-10}
~type-I~& ~$c_\alpha/s_\beta$~ & ~$c_\alpha/s_\beta$~ & ~$c_\alpha/s_\beta$~
& ~$s_\alpha/s_\beta$~ & ~$s_\alpha/s_\beta$~ & ~$s_\alpha/s_\beta$~
& ~$1/\tb$~ & ~$-1/\tb$~ & ~$-1/\tb$~ \\
 \hline
\end{tabular}
\end{center}
 \caption{The Yukawa coupling modifiers in type I 2HDM}
\label{Tab:YukawaFactors}
\end{table} 
The gauge boson couplings to the scalar fields in 2HDM are independent of the Yukawa types. The  couplings of the neutral scalars to a pair of gauge bosons are
\begin{eqnarray}
g_{h_{VV}}=\sba g_{h{VV}}^{SM},\quad g_{H{VV}}=\cba g_{h_{VV}}^{SM},\quad g_{A{VV}}=0 
\label{hVV}
\end{eqnarray} 
where $V= W^\pm,Z$. The $Z$ boson couplings to the neutral scalars are 
\begin{eqnarray}
g_{hAZ_\mu} = \frac{g}{2 c_{\theta_W}}\cba(p_h - p_A)_\mu,\quad g_{HAZ_\mu} = -\frac{g}{2 c_{\theta_W}}\sba(p_H - p_A)_\mu 
\label{hhZ}
\end{eqnarray} 
where $p_\mu$ represent the incoming four momenta of the Higgs bosons, $g$ denotes the $SU(2)_L$ gauge coupling and $\theta_W$ is the Weinberg angle.
Similarly the $W$ boson couplings to the charged Higgs are
\begin{eqnarray}
g_{H^\mp W^\pm h} &=& \mp\frac{ig}{2}\cba(p_h - p_{H^\mp})_\mu,  \nonumber \\
g_{H^\mp W^\pm H} &=& \pm\frac{ig}{2}\sba(p_H - p_{H^\mp})_\mu, \nonumber \\
g_{H^\mp W^\pm A} &=& \frac{g}{2}(p_A - p_{H^\mp})_\mu
\label{hhW}
\end{eqnarray}
with incoming four momenta of the neutral and charged Higgs.
From the Table.~[\ref{Tab:YukawaFactors}], we can see that the fermionic couplings of $H^\pm$ and $A$ are suppressed at large $\tb$ and thus their behaviour is fermiophobic. Another important factor is the alignment limit $\sba \to 1$ \cite{Basler:2017nzu, Branchina:2018qlf} which is mostly favored by the experimental contraints so that all the couplings of $h$ approach to that of the SM. The alignment limit together with large $\tb$ also make the $H$ fermiophobic. This can be seen from the Yukawa coupling modifier of $H$ 
\begin{eqnarray}
\xi^f_H = \frac{s_{\alpha}}{s_{\beta}} = \cba - \frac{\sba}{\tb}. 
\end{eqnarray}
In the limit of $\sba \to 1$ and $\tb >> 1$, $\xi^f_H$ approaches to zero. The fermiophobic behaviour of the BSM Higgs bosons is the most striking characteristics of type I 2HDM.  

For $A$ lighter than half of the mass of the SM Higgs, strong constraint comes from the non-SM $h \to AA$ decay.  
The Higgs trilinear coupling which involves in this decay process is given by 
\begin{eqnarray}
\lambda_{hAA} = \frac{1}{4v^2 s_\beta c_\beta}\Big\lbrace (4M^2 -2m_A^2 -3m_h^2)c_{\alpha+\beta} 
+ (2m_A^2 - m_h^2)c_{\alpha-3\beta}\Big\rbrace
\end{eqnarray}  
and the decay width is given as 
\begin{eqnarray}
\Gamma(h \to AA) = \frac{\lambda_{hAA}^2 v^2}{32 \pi m_h}\sqrt{1 - \frac{4 m_A^2}{m_h^2}}
\end{eqnarray}    
where $M^2=m^2_{12}/s_{\beta}c_{\beta}$. This trilinear coupling $\lambda_{hAA}$ is non vanishing even at the alignment limit $\sba \to 1$. 
\begin{eqnarray}
\lambda_{hAA} &=& \frac{1}{v^2}(2M^2 -2m_A^2 -m_h^2), \quad \forall  \sba \to 1.
\end{eqnarray}

Other trilinear couplings related to the decays of $H$ are $\lambda_{Hhh}$ and $\lambda_{HAA}$, which are given as 
\begin{eqnarray}
\lambda_{Hhh} = \frac{1}{2v^2 s_\beta c_\beta}\cba\Big\lbrace (3M^2 - 2m_h^2 - m_H^2)s_{2\alpha} 
- M^2s_{2\beta} \Big\rbrace,
\end{eqnarray} 
\begin{eqnarray}
\lambda_{HAA} = \frac{1}{4v^2 s_\beta c_\beta}\Big\lbrace (4M^2 - 2m_A^2 - 3m_H^2)s_{\alpha+\beta} 
-(m_H^2 -2m_A^2)s_{\alpha-3\beta} \Big\rbrace
\end{eqnarray}
and in the alignment limit, $\lambda_{Hhh}$  vanishes and $\lambda_{HAA}$ reduces to 
\begin{eqnarray}
\lambda_{HAA} = \frac{2}{v^2t_{2\beta}} (m_H^2 -M^2).
\end{eqnarray} 
For light $A$, the mass splitting between $H^\pm$ and $H$ is highly restricted by the electroweak precision observables, as we will see later, we do not discuss the $HH^+H^-$ coupling. Thus the fermiophobic nature restricts the decay of $H$ only to the $AA$ and $AZ$ modes in the alignment limit.

\section{Electroweak $4b + \ell + {\rlap{\,/}{E}_T}$ state}
\label{4b + W: to probe light $A$}
  In type I 2HDM, the QCD production processes of BSM Higgs bosons are usually suppressed compared to the EW processes due to the fermiophobic behaviour. Not only that, the bosonic decays (both on-shell and off-shell) of $H^\pm \to AW^{(*)}$ and $H\to AA/AZ^{(*)}$ can be the dominant decay modes. Since in our paper we are restricted to light $A$, the branching ratio of $A\to b\overline{b}$ is dominant even though being fermiophobic. Hence, the most promising channel to search for light $A$ is 
\begin{eqnarray}
AAW:p p \to H^\pm A \to (AW)A \to 4b  +  W
\label{sig: AAW}
\end{eqnarray}         
where the signal topology suggests that the prompt $A$ should have higher $p_T$ compared to the $A$ from the decay of $H^\pm$ \cite{Mondal:2023wib}. 
To reduce the QCD multijet background we consider the leptonic decay of $W$ boson. The $4b + \ell +\cancel{\it{E}}_{T}$ would be the final state we are looking for at the LHC. The cross section of the signal at the parton level is 
\begin{eqnarray}
\sigma_{AAW} = \sigma(p p \to H^\pm A)BR(H^\pm \to AW)BR(A\to b\bar{b})^2BR(W \to l \nu) 
\end{eqnarray}
where in the lepton we also include $\tau$. We use a uniform Next-to-Next-to-Leading Order (NNLO) $k$-factor of 1.35 \cite{Bahl:2021str}. 

The dominant background will be the top quark pair production where top quarks decay dominantly to $bW$ mode and the semileptonic and leptonic (including $\tau$) decays of $W$ boson would give at least one lepton. 
 The extra $b$-jets can come from the additional hard jets and from the hadronic decays of $W$ boson faking as $b$-jets. Also for our analysis we generate $t\bar{t}$ background matched up to one parton using the MLM scheme \cite{Alwall:2007fs, Hoeche:2005vzu}. The cross section for the $t\bar{t}+$jets background into fully leptonic and semileptonic states is 458 pb as calculated with Top++ \cite{Czakon:2011xx}. Furthermore, in the case of other backgrounds such as $Wjj$ and $Zjj$, the probability of a QCD jet being misidentified as a $b$-jet is only around $1\%$ \cite{CMS:2012feb}. Hence by mimicing four $b$-jets, the $Wjj$ and $Zjj$ backgrounds can be effectively suppressed and rendered subdominant. Thus for our analysis we consider $t\bar{t}+$jets as the only background.
The cross section is given by
\begin{eqnarray}
\sigma_{BG} = \sigma(p p \to t\bar{t} + jets)BR(t \to b W)^2 BR(W\to \ell \nu)[2-BR(W\to \ell \nu)].
\end{eqnarray}

Before going to the phenomenological study of our signal, we scan the parameter space with two fixed masses of $A$ viz, 50 GeV and 70 GeV. We restrict our study in the scenario of standard mass hierarchy  where we assign the lightest $CP$ even Higgs $h$ as the observed 125 GeV Higgs and $H$ as the heavier $CP$-even Higgs. The model parameters are scanned within the range
\begin{eqnarray}
m_{H^\pm} &:& [80~ - ~350]~ \text{GeV}, \quad m_H : [140~ - ~350]~ \text{GeV}, \quad \tb : [1~ - ~60] \nonumber \\
\sba &:& [0.9~ - ~ 1.0], \quad m_{12}^2 : [0~ - ~350^2]~\text{GeV}^2. 
\label{scan range}
\end{eqnarray}
We randomly generate sample points within the scanning range and apply the theoretical and experimental constraints to obtain the allowed parameter space as listed below. It is worth noting that recent advances in machine learning driven sampling methods \cite{Hammad:2022wpq} can greatly reduce the computational time of such scans.

\begin{figure}[t]
    \centering
    \includegraphics[width=75mm]{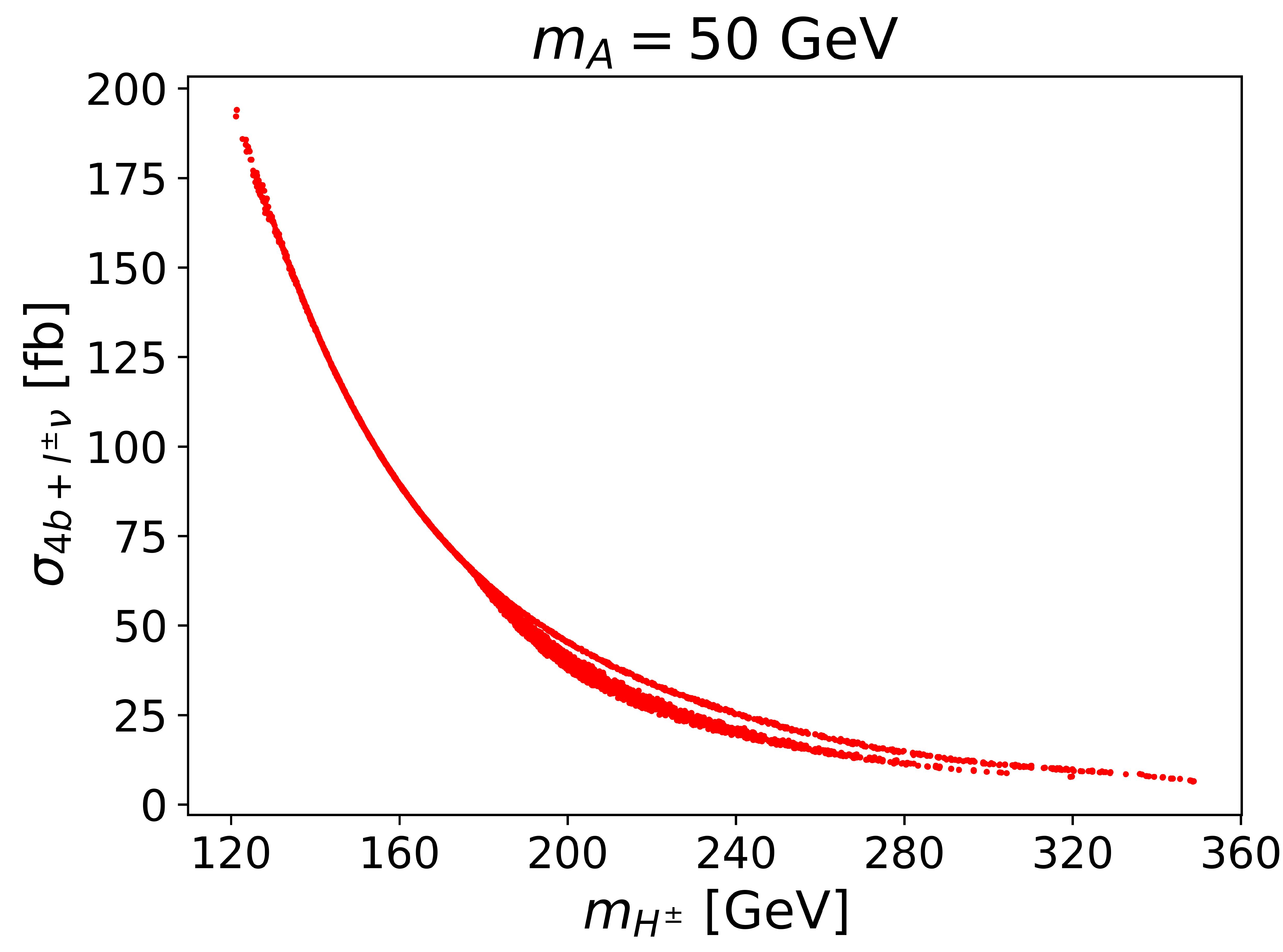}
    \includegraphics[width=75mm]{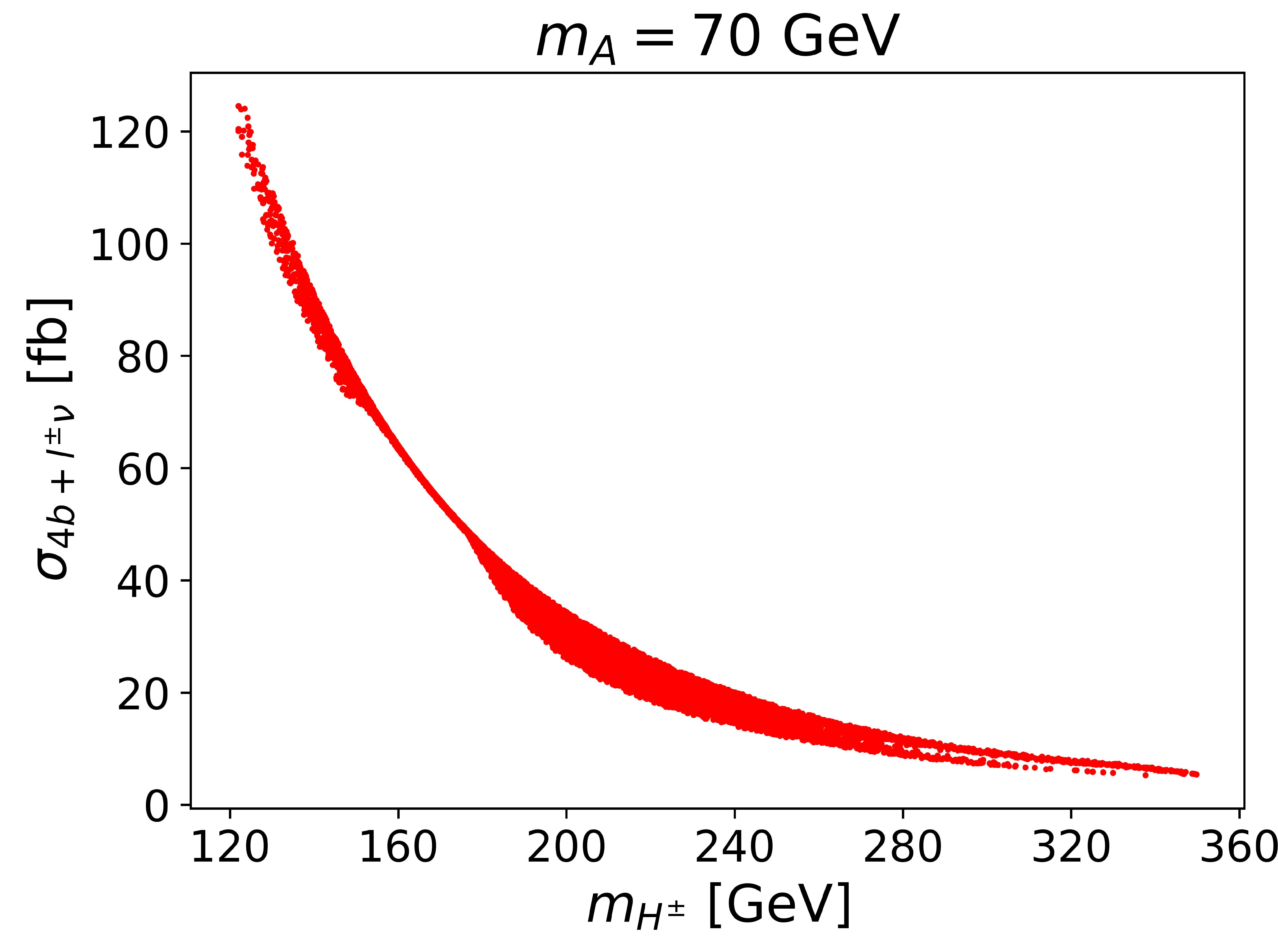}
  \caption{Signal cross sections for the $AAW$ mode with $A\to b\bar{b}$ and $W \to l\nu$. Left: $m_A=50$ GeV and Right: $m_A=70$ GeV.}
  \label{fig: AAW parton level xsec}
\end{figure}

\begin{enumerate}
\item {\bf{Theoretical Constraints:}} The quartic couplings of the scalar potential should be $|\lambda_i| < 4\pi$ \cite{Chang:2015goa} to satisfy the perturbativity condition. The vacuum stability requires the potential to be bounded from below and this gives us the constraints \cite{PhysRevD.18.2574}
\begin{eqnarray}
\lambda_{1,2} > 0, \quad \lambda_3 > -\sqrt{\lambda_1\lambda_2}, \quad \lambda_3 + \lambda_4 -|\lambda_5| > -\sqrt{\lambda_1\lambda_2}.  
\end{eqnarray}   
The tree-level unitarity of the Higgs boson and gauge boson scatterings at high energy as discussed in \cite{Kanemura:1993hm, Akeroyd:2000wc} are also considered. The theoretical constraints are computed using the public code {\tt 2HDMC-1.8.0} \cite{Eriksson:2009ws}. 

\item {\bf Electroweak Precision Observables:} The measurement of the oblique parameters, $S,~T$ and $U$ restricts the mass splitting between the BSM Higgs bosons in the 2HDM, particularly between the charged Higgs $H^\pm$ and the other BSM neutral Higgs bosons ($H,A$). The current best fit results \cite{10.1093/ptep/ptac097} at $95\%$ C.L. are
$S = -0.01 \pm 0.07$ and $T = 0.04 \pm 0.06$ with the correlation $\rho_{ST} = 0.92$ for $U=0$. For the scenario of light $A$, the mass splitting between $H^\pm$ and $H$ gets restricted. We use {\tt 2HDMC} for the computation of the oblique parameters based on the Refs.\cite{Grimus:2007if,Grimus:2008nb}. 

\item {\bf Flavor Physics Constraints:} The $B$-physics observables are calculated using the code {\tt SuperIso-v4.1} \cite{Mahmoudi:2008tp}. The limits on $B \to X_s \gamma$ transition rate \cite{HFLAV:2016hnz, Misiak:2017bgg} excludes $\tb\lesssim 2$ at $95\%$ C.L. in the $m_{H^\pm}-\tb$ plane \cite{Misiak:2017bgg, Sanyal:2019xcp} for type I 2HDM, thus reflecting the fermiophobic behaviour of $H^\pm$ with respect to $\tb$.

\item {\bf Collider Constraints:} The exclusion limits for the direct Higgs boson searches at LEP,
Tevatron and LHC at $95\%$ C.L. are imposed by using the public code {\tt HiggsBounds-v5.10.2} \cite{Bechtle:2020pkv}. Along with the direct searches, we also check the consistency of the Higgs precision measurements using the code {\tt HiggsSignals-v2.6.2} \cite{Bechtle:2020uwn}. We find the allowed parameter points at $95\%$ C.L. with respect to the best fit point in two dimensional parameter spaces, which corresponds to $\Delta\chi^2_{\text{HS}} = \chi^2_{\text{HS}} - \chi^2_{\text{HS,min}} \lesssim 6$. 
The $\chi^2_{\text{HS,min}}$ for the best fit points of $m_A$ = 50 GeV and 70 GeV cases are approximately 92 and 90 respectively.       
\end{enumerate}

After imposing all the constraints we obtain the allowed parameter space, for which we compute the parton level cross sections for the signal [\ref{sig: AAW}], considering the leptonic decay of $W$ boson as shown in Fig.[\ref{fig: AAW parton level xsec}]. For $m_A = 50$ GeV, $h\to AA$ decay is allowed.
The constraints from the SM Higgs exotic decay includes $h \to AA \to bbbb$ ATLAS \cite{ATLAS:2018pvw}, $h \to AA \to bb \tau \tau$ CMS \cite{CMS:2018zvv}, $h \to AA \to bb \mu \mu$ ATLAS \cite{ATLAS:2018emt}  CMS \cite{CMS:2017dmg, CMS:2018nsh}, $h \to AA \to \tau\tau\tau\tau$ CMS \cite{CMS:2017dmg}, $h \to AA \to \tau\tau\mu\mu$ CMS \cite{CMS:2017dmg, CMS:2018qvj}. Along with these, strong constraints also come from the precisely measured Higgs decay width \cite{CMS:2022dwd, CMS:2019ekd, CMS:2022ley, ATLAS:2023dnm}, thus restricting the Higgs trilinear coupling, $|\lambda_{hAA}|\lesssim 0.013$. To study the discovery prospects of the signal at the LHC, we proceed our analysis with some benchmark points (BPs) as given in Table.~[\ref{Benchmark Points}].


\begin{table}[t]
\setlength{\tabcolsep}{2pt}
\centering
\begin{tabular}{ccccccc}
\toprule
Signals~~ & ~~$m_A$ [GeV]~~ & ~~$m_{H^\pm}$ [GeV]~~ & ~~$m_H$ [GeV]~~ & ~~$s_{\beta-\alpha}$~~ & ~~$m^2_{12}$ [GeV$^2$]~~ & ~~$t_\beta$ \\
\midrule
BP1  & 50 & 142.811 & 141.438 & 0.95771 & 1209.72 & 15.9983 \\
BP2  & 50 & 184.916 & 161.629 & 0.95998 & 2333.38 & 10.7302 \\
BP3  & 50 & 225.747 & 208.539 & 0.95998 & 4724.73 & 8.4401 \\
\midrule
BP4  & 70 & 152.41 & 159.024 & 0.98344 & 3123.09 & 6.05755 \\
BP5  & 70 & 190.812 & 177.972 & 0.98955 & 3651.57 & 8.09766 \\
BP6  & 70 & 236.081 & 219.12 & 0.96523 & 6073.61 & 7.04902 \\
\bottomrule
\end{tabular}
 \caption{ Benchmark points with $m_A = 50$ and 70 GeV.}
 \label{Benchmark Points}
\end{table}

\subsection{$\chi^2$  Method}

We see from Fig.[\ref{fig: AAW parton level xsec}], the cross sections of the signal is only of the order of 100 fb, however, the $t\bar{t}+$jets background cross section (458 pb) is significantly higher than the signal. Thus simple cut based analysis based on $n_b,~p_T,~\Delta R, ~{\rlap{\,/}{E}_T}$ variables are not sufficient to probe any excess of the signal over the background. Thus we need to construct suitable variable or discriminator based on the signal topology and signal hypothesis e.g. masses of the new physics which are $m_{H^\pm}$ and $m_A$ in our case. In this work we construct a $\chi^2$ which is given by 
\begin{eqnarray}
\chi^2 = \Big( \frac{m_{bb} - m_A}{\sigma_{m_A}} \Big)^2 + \Big( \frac{m_{bbl\nu} - m_{H^\pm}}{\sigma_{m_{H^\pm}}} \Big)^2.
\label{chi square}
\end{eqnarray}
We explain the $\chi^2$ method in the following way:

\begin{enumerate}
\item {\bf $b$-jet pairing algorithm:} The $AAW$ signal would give at least four resolved $b$-jets and one lepton final state. Two $b$-jet pairs are constructed out of the four leading $b$-jets. There are three possible combinations to make $b$-jet pairs. We use subscripts 1 and 2 to refer the $b$-jet pairs. A jet pairing algorithm is used to choose one of the three possible combinations. We label the $b$-jets with the subscript $a,b,c$ and $d$ and the three combinations would be (1,2; 3,4), (1,3; 2,4) and (1,4; 2,3). The pairing algorithm considers the combination which minimises \cite{CMS:2018mts, CMS:2022usq}
\begin{eqnarray}
\Delta R = |(\Delta R_1 - 0.8)| + |(\Delta R_2 - 0.8)| 
\end{eqnarray}  
where $\Delta R_1$ and $\Delta R_2$ for a particular combination are given by 
\begin{eqnarray}
\Delta R_1 &=& \sqrt{(\eta_a - \eta_b)^2 + (\phi_a - \phi_b)^2}, \nonumber \\
\Delta R_2 &=& \sqrt{(\eta_c - \eta_d)^2 + (\phi_c - \phi_d)^2}.
\end{eqnarray}
The pairing algorithm is motivated by the idea that the $b$-jets from the pseudoscalars would be closer together compared to the uncorrelated $b$-jets. The offset of 0.8 is used to reduce the pairings where the $b$-jets ovelap in the $\eta-\phi$ space.   

\item {\bf Calculating $p_{z_\nu}$ of neutrino:} In the $AAW$ signal, we consider the leptonic decay of $W$ boson. However, since there is no reconstructed object at the detector that corresponds to the neutrino, only the transverse component of the momentum can be inferred from the conservation of momentum: $\overrightarrow{p}_{T_\nu} = -\sum_i \overrightarrow{p}_{T_i}$ ($i$ includes the observed particles). \\
The $z$-component can be computed using the on-shell condition
\begin{eqnarray}
m_{W}^2 = (E_\ell + E_\nu)^2 -(\overrightarrow{p}_\ell + \overrightarrow{p}_\nu)^2.
\end{eqnarray}
Rewriting this in terms of the $x,y,z$ components of the neutrino momentum, we get a quadratic equation 
\begin{eqnarray}
A p_{z_\nu}^2 + B p_{z_\nu} + C =0
\label{pz neutrino}
\end{eqnarray}
where the coefficients are 
\begin{eqnarray}
A &=& 4(E_\ell^2 - p^2_{z_\ell}), \nonumber \\
B &=& -4ap_{z_\ell}, \nonumber \\
C &=& 4E_\ell^2 (p^2_{x_\nu} + p^2_{y_\nu}) -a^2
\end{eqnarray}
and $a = m_W^2 - m_\ell^2 + 2p_{x_{\ell}}p_{x_{\nu}} + 2p_{y_{\ell}}p_{y_{\nu}}$. Solving Eq.[\ref{pz neutrino}] we get 
\begin{eqnarray}
p_{z_\nu} = \frac{-B \pm \sqrt{B^2 - 4AC}}{2A}
\end{eqnarray}
and in the case of imaginary root, the real component has to be considered. The estimation of the $z$-component of the neutrino momentum is possible when the missing transverse momentum corresponds to one neutrino. If the process consists of multiple neutrinos the above method does not hold true. Thus the method of obtaining $p_{z_\nu}$ is appropriate strictly for $W\to e\nu_e$ and $W\to\mu \nu_\mu$ but not for $W \to \tau\nu_\tau \to e 2\nu_\tau \nu_e$ and $W \to \tau\nu_\tau \to \mu 2\nu_\tau \nu_\mu$. The direct production of $e/\mu$ via $W$ boson decay amounts to $22\%$ of the total $W$ boson decay width. Whereas, the indirect production of $e/\mu$ via $W \to\tau\nu_\tau$ decay is only $4\%$ of the $W$ boson decay and therefore contributes insignificantly. Hence, we consider $e/\mu$ production via $W \to \tau \nu_{\tau}$ in conjunction with $W \to e \nu_{e}$ and $W \to \mu \nu_{\mu}$ to implement the above procedure. 

\item {\bf Mass resolutions:} The mass resolution refers to the expected uncertainty in the measurement of the masses of the BSM particles at the detector. The signal $AAW$ has two pseudoscalars, one of the pseudoscalar comes from the decay of $H^\pm$ and the other is the prompt pseudoscalar. To a good approximation we can assume that the leading $b$-jet comes from the prompt $A$. We can reconstruct the mass of $A$ from the invariant mass distribution of the $b$-jet pair which contains the leading $b$-jet. The other $b$-jet pair together with the lepton and the neutrino reconstructs the mass of $H^\pm$. However, since the neutrino cannot be observed at the detector, we cannot estimate the $4$-momentum of the neutrino\footnote{The above method of $p_{z_\nu}$ gives two possible solutions instead of exact solution required to reconstruct the $4$-momentum.}. Hence the neutrino used for the truth reconstruction of $H^\pm$ is the generator level neutrino \cite{CMS:2017ixp}. The mass resolutions $\sigma_{m_A}$ and $\sigma_{m_{H^\pm}}$ are estimated by obtaining the full width at half maximum (FWHM) of the invariant mass distributions of $m_{bb}$ and $m_{bbl\nu}$. Fig.[\ref{fig: width}] gives the mass resolutions of both $A$ and $H^\pm$ for BP1 and BP6 and Table.~[\ref{Benchmark points width}] gives the mass resolutions of the selected BPs\footnote{The mass resolutions obtained from the widths of the invariant mass distributions depend on the detector sensitivity, which in our case is consistent when we compare with the mass resolutions of $W$ boson and top quark mentioned in Ref.\cite{CMS:2017ixp}.}. These mass resolutions are used as inputs to $\chi^2$ in Eq.[\ref{chi square}].  

\begin{figure}[t]
    \centering
    \includegraphics[width=75mm]{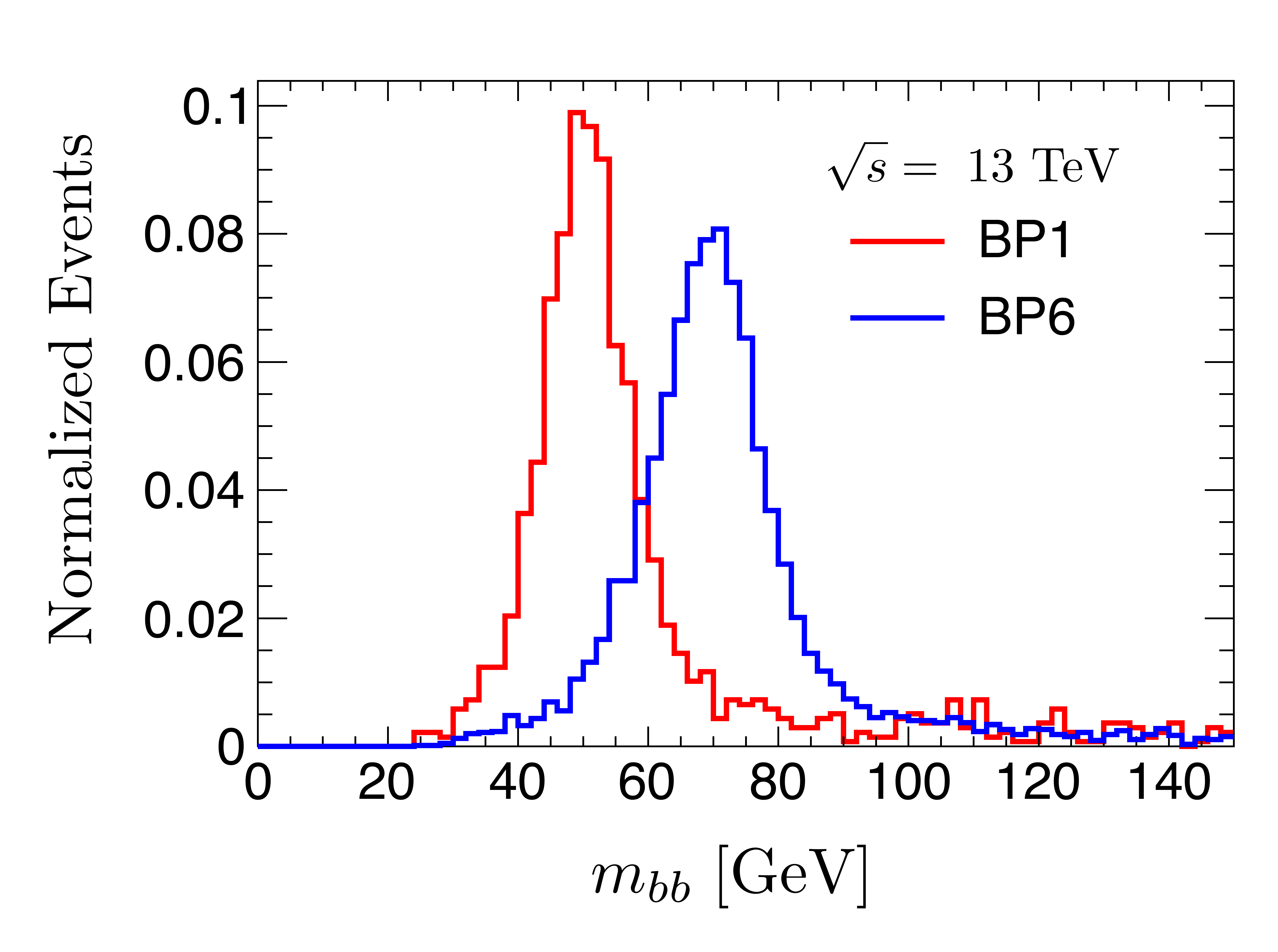}
    \includegraphics[width=75mm]{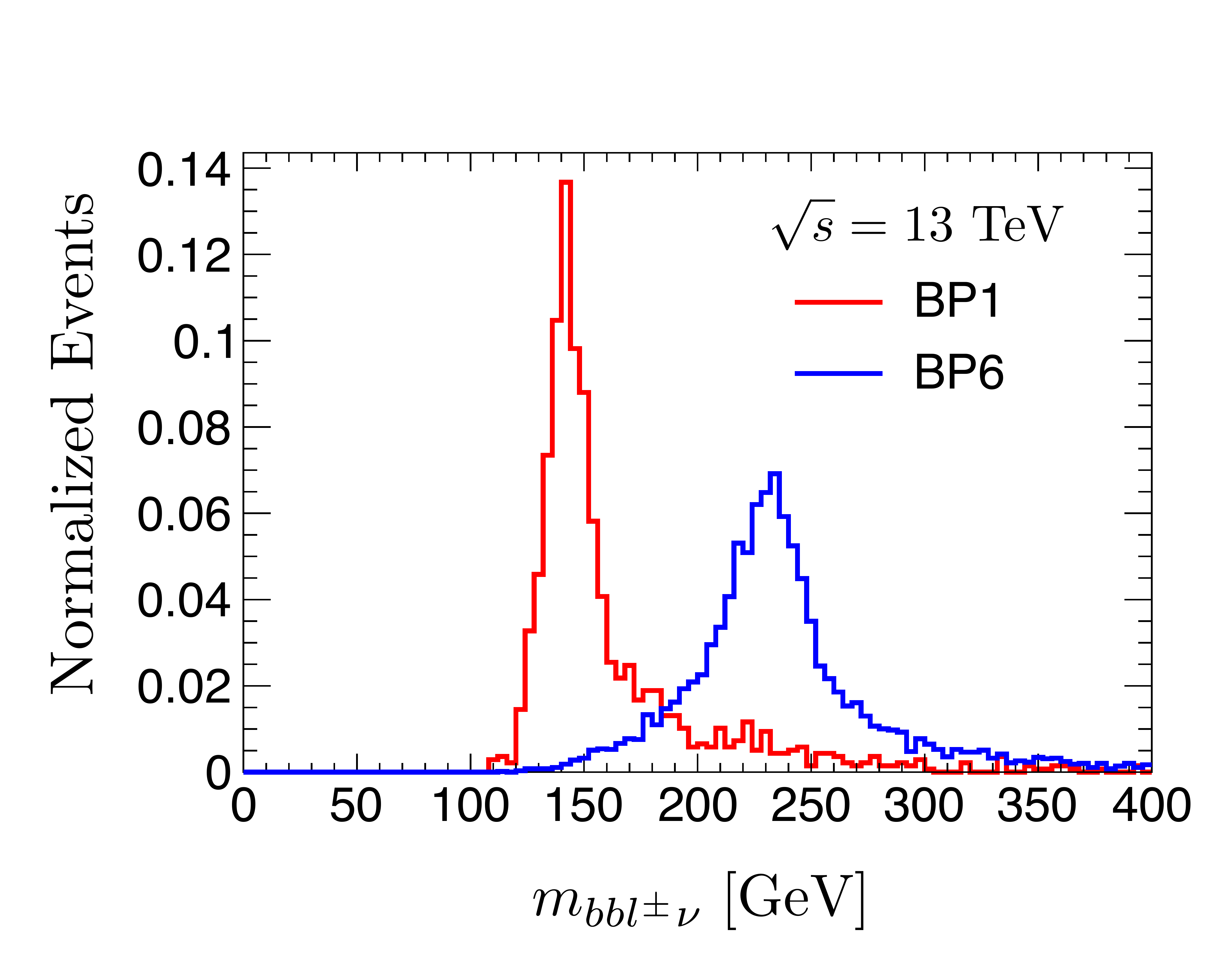}
  \caption{$m_{bb}$ and $m_{bbl\nu}$ distributions normalized to one event for the BP1 and BP6. The FWHM gives the mass resolutions, $\sigma_{m_A}$ and $\sigma_{m_{H^\pm}}$ respectively.}
  \label{fig: width}
\end{figure}

\begin{table}[t]
\centering
\begin{tabular}{ccccc}
\toprule
Signals~~ & ~~$m_{A}$ [GeV]~~ & ~~$m_{H^\pm}$ [GeV]~~ & ~~$\sigma_{m_A}$ [GeV]~~ & ~~$\sigma_{m_{H^\pm}}$ [GeV] \\
\midrule
BP1  & 50 & 142.811 & 13.97 & 19.92 \\
BP2  & 50 & 184.916 & 12.13 & 35.81 \\
BP3  & 50 & 225.747 & 12.14 & 36.15  \\
\midrule
BP4  & 70 & 152.41 & 17.92 & 28.19 \\
BP5  & 70 & 190.812 & 18.00 & 28.00  \\
BP6  & 70 & 236.081 & 17.92 & 40.23  \\
\bottomrule
\end{tabular}
\caption{Mass resolutions of $A$ and $H^\pm$ for the selected BPs.}
\label{Benchmark points width}
\end{table}

\item {\bf $\chi^2$ per event:} The $\chi^2$ based on the signal hypothesis will be used to obtain an excess of $4b+\ell + {\rlap{\,/}{E}_T}$ over the background. In each event (signal and background), we compute the possible combinations of $\chi^2$s and pick the one which is minimum as the $\chi^2$ of the event. The combinations are done based on the two $b$-jet pairs  and the possible solutions of $p_{z_\nu}$. Since the $\chi^2$ is constructed based on the signal, the $\chi^2$ for the signal is expected to be very small compared to the background. We can see this feature in Fig.[\ref{fig: chi_square}] for BP1 and BP6 where the signal is concentrated on small values of $\chi^2$ and falls rapidly with $\chi^2$ and for the background the distribution is very broad. Here as a signal we consider the $4b+\ell + {\rlap{\,/}{E}_T}$ final state only through $AAW$ mode and $t\bar{t}+$jets as the background. Also the distributions are obtained after imposing the basic selection cuts which we will discuss in Sec.[\ref{Signal-background analysis}].   

\begin{figure}[t]
    \centering
    \includegraphics[width=75mm]{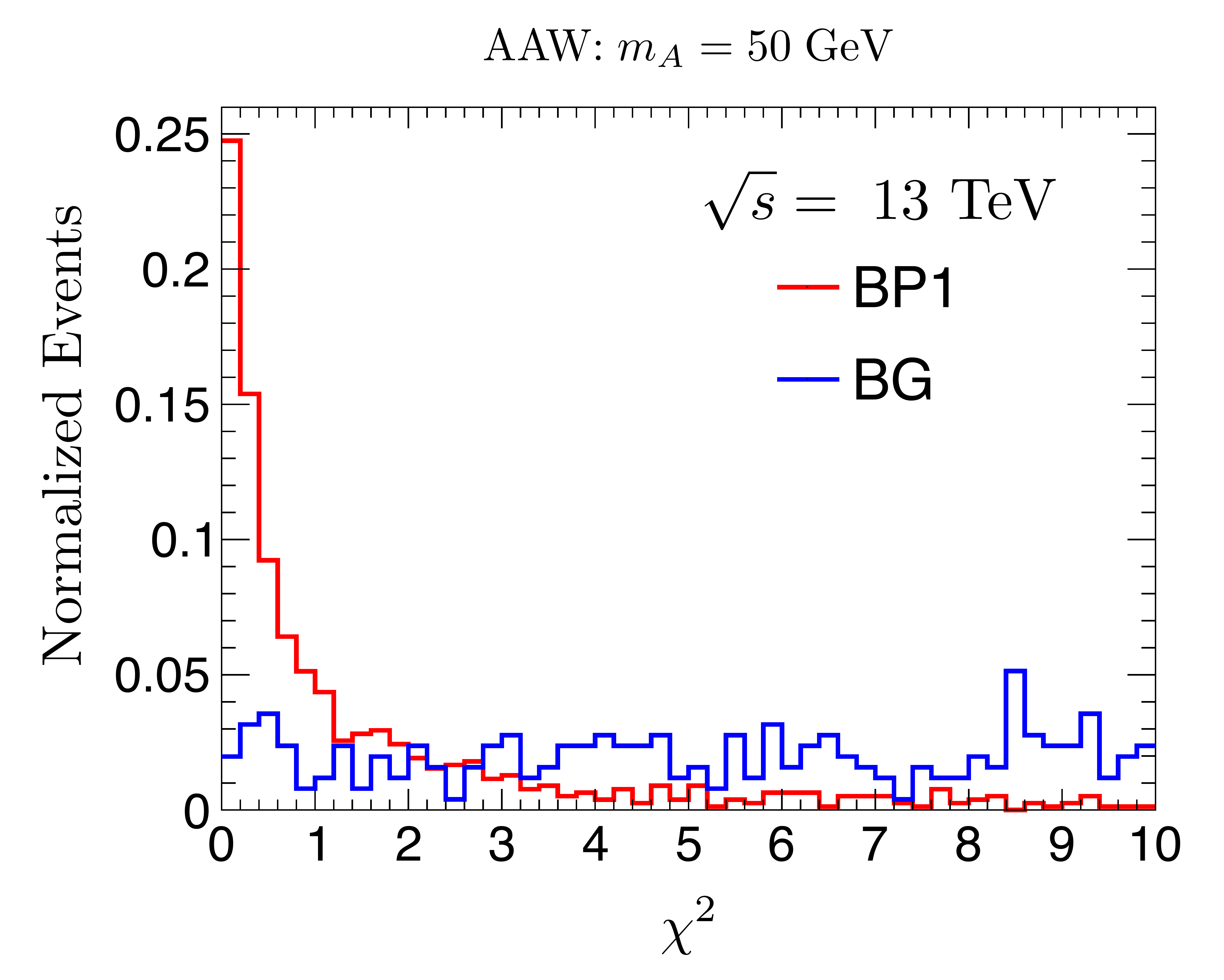}
    \includegraphics[width=75mm]{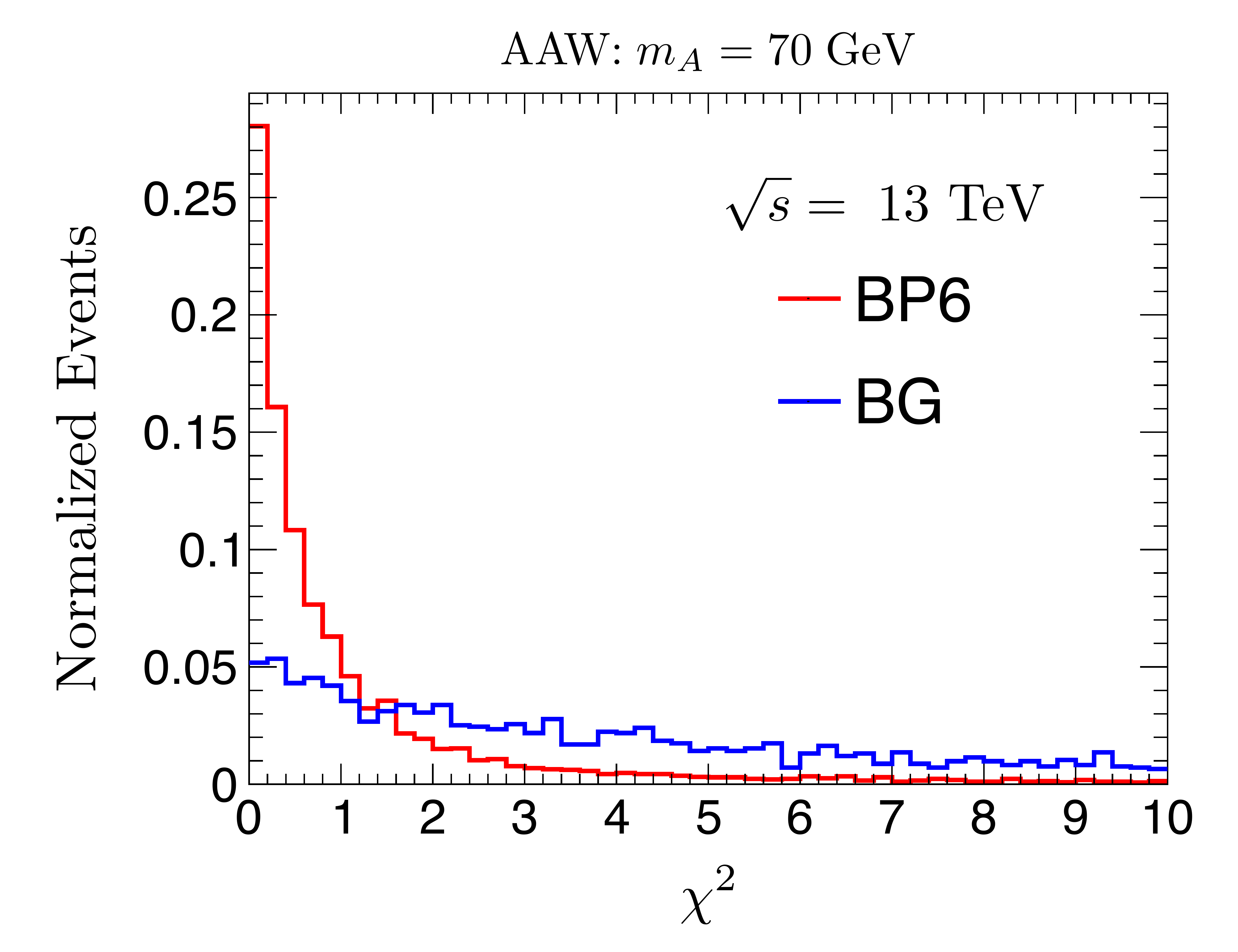}
  \caption{$\chi^2$ distributions normalized to one event for the signal (via AAW mode) and background.  }
  \label{fig: chi_square}
\end{figure}
 
\end{enumerate}

\subsection{Signal-background analysis}  
\label{Signal-background analysis}
In this section, we perform the signal-background analysis for the final state $4b+\ell + {\rlap{\,/}{E}_T}$ at the LHC through the detector simulation. The signal considered so far is the $AAW$ mode as given in Eq.[\ref{sig: AAW}]. There can be additional contributions to $4b+\ell + {\rlap{\,/}{E}_T}$ from other EW processes as given below: 
\begin{eqnarray}
AAAW&:& p p \to H^\pm H \to (A W)(AA) \to 4b + \ell + {\rlap{\,/}{E}_T} + X, \nonumber \\
AAZW&:& p p \to H^\pm H \to (A W)(AZ) \to 4b + \ell + {\rlap{\,/}{E}_T} + X, \nonumber \\
AAWW&:& p p \to H^+ H^- \to (A W)(A W) \to 4b + \ell + {\rlap{\,/}{E}_T} + X
\label{subdiminant signals}
\end{eqnarray}
however, their contributions at the parton or generator level is subdominant compared to the $AAW$ mode. 
Here $X$ can be any jets (including $b$-jets) and/or leptons. Note that in the $AAZW$ process we allow all possible decay modes of $Z$ bososn. Since the QCD corrections are only through the initial states and it would be same for the charged current and neutral current, the same $k$-factor of 1.35 can be imposed to all the EW processes. The cross sections at the parton level for 13 TeV LHC are given by
\begin{eqnarray}
\sigma_{AAAW} &=& \sigma (pp \to H^\pm H)BR(H^\pm \to AW)BR(H \to AA)BR(A \to b\bar{b})^2 \nn \\
&&BR(W\to \ell \nu)[3-2BR(A\to b\bar{b})], \nn \\
\sigma_{AAZW} &=& \sigma (pp \to H^\pm H)BR(H^\pm \to AW)BR(H \to AZ)BR(A \to b\bar{b})^2 \nn \\
&&BR(W\to \ell \nu), \nn\\
\sigma_{AAWW} &=& \sigma (pp \to H^\pm H^\pm)BR(H^\pm \to AW)^2BR(A \to b\bar{b})^2BR(W\to \ell \nu) \nn \\
&&[2-BR(W \to \ell\nu)]. 
\end{eqnarray}
The cross sections are based on at least four $b$-quarks and at least one lepton. The leptonic decay of $W$ boson includes $\tau$ as well.
For Monte Carlo event generation, the type I 2HDM model is first implemented in {\tt FeynRules-2.3} \cite{Alloul:2013bka}. Then the event generation for signal and background are done using {\tt MadGraph5\_aMC@NLO} \cite{Alwall:2011uj} with {\tt NNPDF31\_lo\_as\_118} parton distribution functions set \cite{NNPDF:2017mvq}. We used {\tt PYTHIA-8.2} \cite{Sjostrand:2014zea} for parton showering and hadronization. For detector simulation we used {\tt Delphes-3.4.2} \cite{deFavereau:2013fsa}. We use anti-kt algorithm \cite{Cacciari:2008gp} with radius parameter $R =0.4$ and $p_T (j)>20$ GeV for jet reconstruction. We also followed the default $b$-jet (mis-)tagging efficiencies as given in the {\tt Delphes} CMS card based on Ref.\cite{CMS:2012feb}. After the generation of signal and background events, we impose the following basic selection cuts:
\begin{enumerate}
\item We select events with at least four $b$-jets and at least one lepton ($e,\mu$).
\item  The $b$-jets and lepton(s) are required to satisfy the criteria\\
\begin{center}
$p_T^b > 20$ GeV, \quad $p_T^\ell > 10 $ GeV, \quad $|\eta^{b,\ell}| < 2.5$.
\end{center}      
\item We impose a nominal cut on the missing transverse energy (MET): ${\rlap{\,/}{E}_T} > 10$ GeV.
\end{enumerate}

\begin{figure}[t]
    \centering
    \includegraphics[width=75mm]{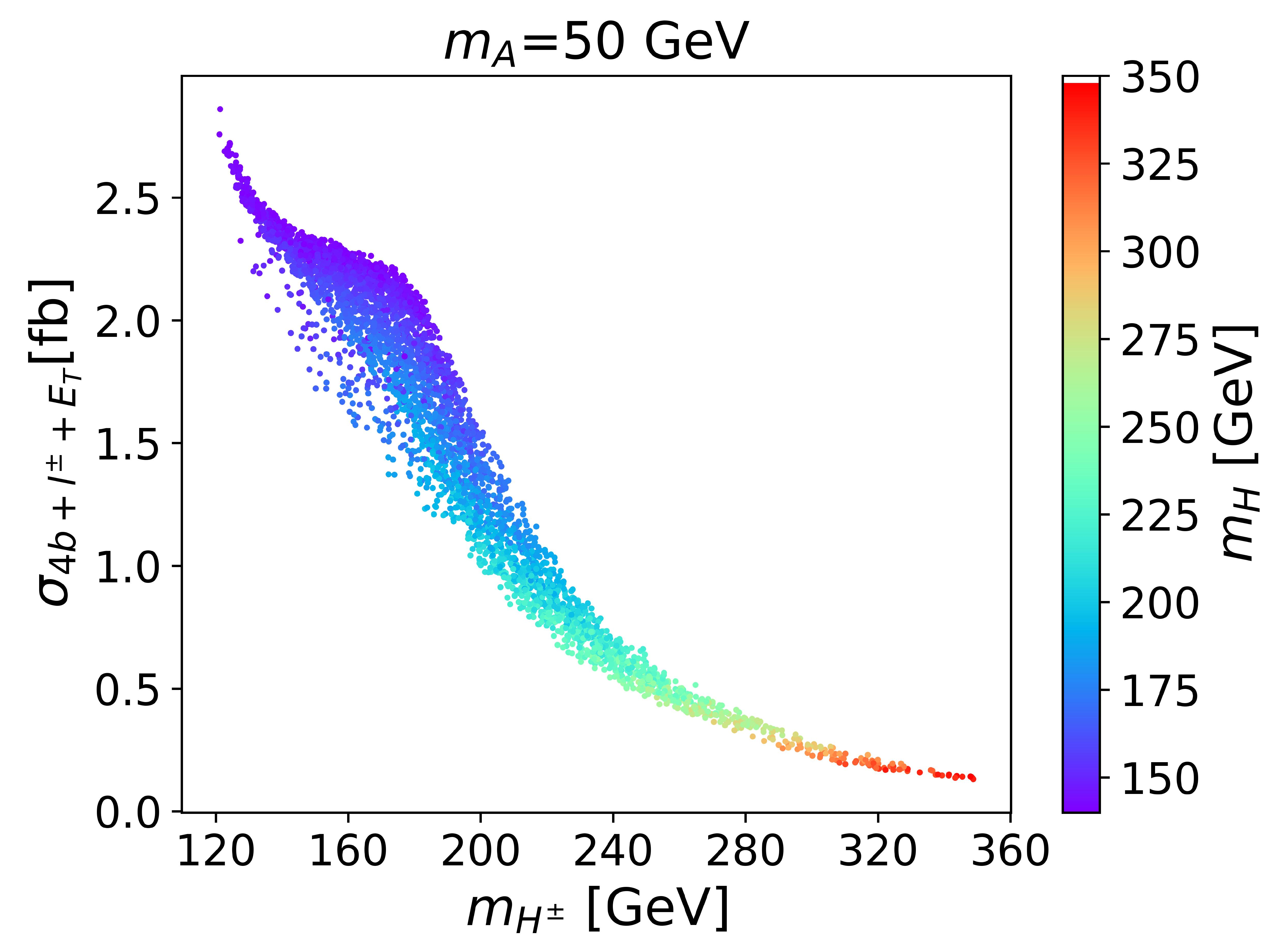}
    \includegraphics[width=75mm]{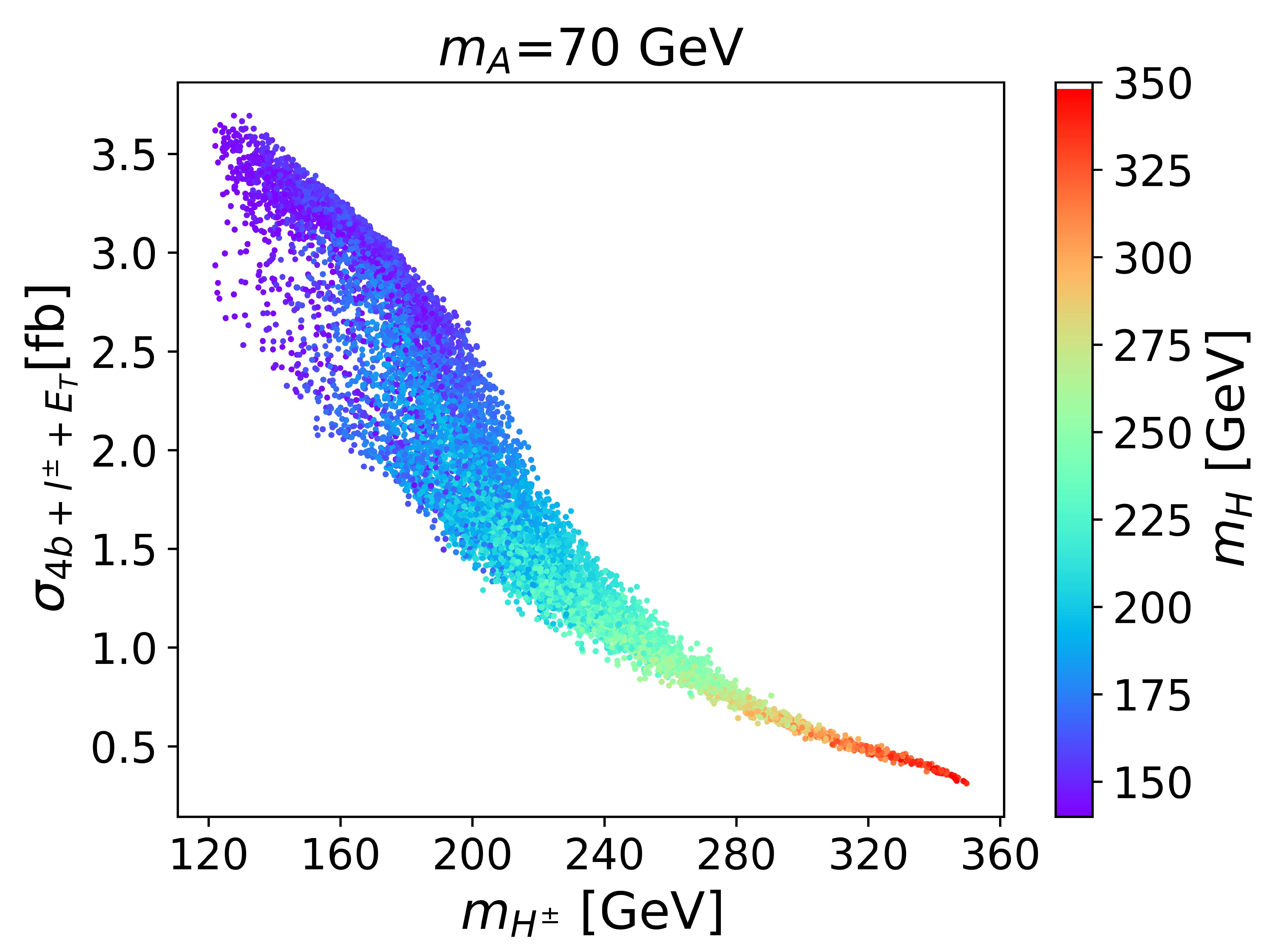}
  \caption{Total signal cross sections at the detector level with basic selection cuts. Left: $m_A = 50$ GeV and Right: $m_A = 70$ GeV.}
  \label{fig: signal cross section with basic cuts}
\end{figure}

The total signal cross sections for the final state $4b + \ell + {\rlap{\,/}{E}_T}$ after imposing the basic selection cuts are shown in Fig.[\ref{fig: signal cross section with basic cuts}]. There is a spread in the cross section if we compare with Fig.[\ref{fig: AAW parton level xsec}] because of the contributions coming from the subdominant channels which depend on other model parameters like $M^2$ through $H \to AA$ decay and $s_{\beta -\alpha}$ through $H^\pm H$ production. We also see that the mass splitting between $H^\pm$ and $H$ is restricted due to the constraints from the EWPOs. The tagging of four $b$-jets at the detector reduces the cross sections of the signal to few fb whereas the cross section of the background after the basic selection is 772.2 fb which is still significantly higher than the proposed signal. \par
 
Therefore we resort to additional selection cuts as given below:
\begin{enumerate}
\item We impose a strong selection cut on $\chi^2$, which is
\begin{eqnarray}
 \chi^2 < 1
\end{eqnarray} 
 to eliminate the background significantly. Even though we have additional sources of signal via EW processes Eq.[\ref{subdiminant signals}] to the $4b + \ell + {\rlap{\,/}{E}_T}$ final state. We impose the same $\chi^2$ as given in Eq.[\ref{chi square}], since their contribution at the generator level is subdominant.

\item After that, we require that the two pairs of $b$-jets which are obtained by the pairing algorithm should satisfy the asymmetry cut \cite{CMS:2022usq}
\begin{eqnarray}
\alpha = \frac{|m_1 - m_2|}{m_1 + m_2} < 0.1
\end{eqnarray}
where $m_1$ and $m_2$ are the invariant masses of the two $b$-jet pairs. The asymmetry cut ensures that the two $b$-jet pairs are from identical sources ($AA$ pair in our signal). The asymmetry cut suits best to the dominant $AAW$ mode, however just like the $\chi^2$ cut we apply the same asymmetry cut to the other signals as they contribute subdominantly at the generator level.  

\item Finally we impose the $\eta$ separation of the $b$-jet pairs to satisfy 
\begin{eqnarray}
\Delta \eta = |\eta_1 - \eta_2| < 1.1
\end{eqnarray}
to reduce the background contributions via $t$-channel process \cite{CMS:2022usq}. Here $\eta_1$ and $\eta_2$ are defined as 
\begin{eqnarray}
\eta_1 = \frac{\eta_{a} + \eta_{b}}{2}, \quad \eta_2 = \frac{\eta_{c} + \eta_{d}}{2}
\end{eqnarray}
where $(a,b)$ and $(c,d)$ are the two $b$-jet pairs obtained by the pairing algorithm.
\end{enumerate}

\begin{table}[ht]
\setlength{\tabcolsep}{2pt}
\begin{center}
\begin{tabular}{|c|c|c|c|c|c|c|c|c|c|c|}
\hline \tabincell{c}{~~Cut~~ \\~ Flow~~} & \tabincell{c}{Signals} & \tabincell{c}{~~Modes~~} &\tabincell{c}{~~Parton~~ \\ ~Level~}  &  \tabincell{c}{~~Basic~~ \\ ~Cut~~} & \tabincell{c}{~~$\chi^2$<1~~} & \tabincell{c}{~~$\alpha$ < 0.1~~}  &\tabincell{c}{~~$\Delta \eta$ < 1.1~~} & \tabincell{c}{~~Significance~~} \\
\hline \multicolumn{1}{|c|}{\multirow{15}{*}{\makecell{$m_A=$\\50 GeV}}} & \multirow{5}{*}{BP1} & AAW & $125.69$ & $0.504$ & $0.239$ & $0.164$ & $0.132$ & \multirow{5}{*}{13.16}\\ 
\multicolumn{1}{|c|}{} &  & AAAW & $37.59$ & $1.174$ & $0.143$ & $0.081$ & $0.069$ &\\
\multicolumn{1}{|c|}{} &  & AAZW & $5.01$ & $0.048$ & $0.005$ & $0.003$ & $0.002$ &\\ 
\multicolumn{1}{|c|}{} &  & AAWW & $35.26$ & $0.317$ & $0.043$ & $0.027$ & $0.024$ &\\
\multicolumn{1}{|c|}{} &   & BG &$458000$ & $772.2$ & $5.04$ & $1.28$ & $0.82$ &\\ 
\cline{2-9} \multicolumn{1}{|c|}{} & \multirow{5}{*}{BP2} & AAW &$57.62$ & $0.440$ & $0.268$ & $0.177$ & $0.144$ & \multirow{5}{*}{5.91}\\  
\multicolumn{1}{|c|}{} &  & AAAW &$1.36$ & $0.074$ & $0.013$ & $0.005$ & $0.004$ &\\
\multicolumn{1}{|c|}{} &  & AAZW &$14.53$ & $0.303$ & $0.051$ & $0.025$ & $0.021$ &\\  
\multicolumn{1}{|c|}{} &  & AAWW &$13.54$ & $0.221$ & $0.055$ & $0.033$ & $0.027$ &\\
\multicolumn{1}{|c|}{} &   & BG &$458000$ & $772.2$ & $28.85$ & $6.41$ & $3.21$ &\\
\cline{2-9} \multicolumn{1}{|c|}{} & \multirow{5}{*}{BP3} & AAW &$30.72$ & $0.298$ & $0.181$ & $0.122$ & $0.096$ & \multirow{5}{*}{3.13}\\
\multicolumn{1}{|c|}{} &  & AAAW &$1.13$ & $0.077$ & $0.012$ & $0.004$ & $0.003$ &\\
\multicolumn{1}{|c|}{} &  & AAZW &$5.45$ & $0.155$ & $0.029$ & $0.014$ & $0.011$ &\\
\multicolumn{1}{|c|}{} &  & AAWW &$6.51$ & $0.115$ & $0.032$ & $0.019$ & $0.013$ &\\
\multicolumn{1}{|c|}{} &   & BG &$458000$ & $772.2$ & $58.62$ & $9.39$ & $4.58$ &\\ 
\hline \multicolumn{1}{|c|}{\multirow{15}{*}{\makecell{$m_A=$\\70 GeV}}} & \multirow{5}{*}{BP4} & AAW &$72.73$ & $0.665$ & $0.372$ & $0.230$ & $0.194$ & \multirow{5}{*}{11.20}\\ 
\multicolumn{1}{|c|}{} &  & AAAW &$33.18$ & $1.306$ & $0.282$ & $0.137$ & $0.117$ &\\ 
\multicolumn{1}{|c|}{} &  & AAZW &$0.002$ & $-$ & $-$ & $-$ & $-$ &\\ 
\multicolumn{1}{|c|}{} &  & AAWW &$25.00$ & $0.379$ & $0.078$ & $0.042$ & $0.035$ &\\ 
\multicolumn{1}{|c|}{} &   & BG & $458000$ & $772.2$ & $18.23$ & $4.40$ & $2.75$ & \\ 
\cline{2-9} \multicolumn{1}{|c|}{} & \multirow{5}{*}{BP5} & AAW &$38.84$ & $0.592$ & $0.321$ & $0.213$ & $0.169$ & \multirow{5}{*}{4.93}\\ 
\multicolumn{1}{|c|}{} &  & AAAW &$5.20$ & $0.363$ & $0.077$ & $0.039$ & $0.032$ &\\ 
\multicolumn{1}{|c|}{} &  & AAZW &$8.13$ & $0.244$ & $0.050$ & $0.027$ & $0.022$ &\\ 
\multicolumn{1}{|c|}{} &  & AAWW &$10.79$ & $0.284$ & $0.070$ & $0.042$ & $0.033$ &\\ 
\multicolumn{1}{|c|}{} &   & BG &$458000$ & $772.2$ & $43.05$ & $13.97$ & $8.02$ &\\ \cline{2-9} \multicolumn{1}{|c|}{} & \multirow{5}{*}{BP6} & AAW &$20.20$ & $0.394$ & $0.244$ & $0.161$ & $0.127$ & \multirow{5}{*}{2.92}\\ 
\multicolumn{1}{|c|}{} &  & AAAW &$1.32$ & $0.139$ & $0.032$ & $0.014$ & $0.011$ &\\ 
\multicolumn{1}{|c|}{} &  & AAZW &$3.75$ & $0.165$ & $0.044$ & $0.022$ & $0.017$ &\\ 
\multicolumn{1}{|c|}{} &  & AAWW &$4.79$ & $0.164$ & $0.056$ & $0.033$ & $0.026$ &\\ 
\multicolumn{1}{|c|}{} &   & BG &$458000$ & $772.2$ & $98.93$ & $21.76$ & $11.45$ &\\ \hline     
\end{tabular}
\caption{The cut-flow table of the cross sections (in units of fb) for the signal and background with $m_A=$ 50 GeV and 70 GeV at the 13 TeV LHC. Here "$-$" implies that the cross sections are insignificant. The significances are estimated for $3000$ fb$^{-1}$ luminosity.}
\label{cutflow}
\end{center}

\end{table}

We can now estimate discovery prospects of the $4b + \ell + {\rlap{\,/}{E}_T}$ signal at the 13 TeV LHC for $3000$ fb$^{-1}$ luminosity by using the significance estimator given by \cite{Cowan:2010js}
\begin{eqnarray}
S = \sqrt{2\Big[ (N_s + N_b) \log \Big( 1 + \frac{N_s}{N_b}\Big) -N_s \Big]}
\label{Significance}
\end{eqnarray}
where $N_s$ and $N_b$ are the number of signal and background events obtained after imposing all the selection cuts. To see the discovery reach of the signal we use the BPs as given in Table.~[\ref{Benchmark Points}]. The $\chi^2$ method would hold as long as the $W$ boson is produced on-shell. Hence in our analysis for the computation of significance, we only consider the BPs for which $H^\pm \to AW$ is on-shell. In Table.~[\ref{cutflow}], the cross sections for the signal (dominant as well as subdominant contributions) and background are given with the selection cuts imposed sequentially. The basic cut depends only on the tagging efficiencies of $4b$-jets. Since the $AAAW$ mode has an additional $A$ compared to the $AAW$ mode, it allows for the possibility of additional $b$-jets. As a result, the $4b$ tagging efficiency is higher for the $AAAW$ mode, leading to its dominance over the $AAW$ mode at the basic cut for some BPs, despite lower cross sections at the parton level. The $\chi^2$ discriminator is based on the $AAW$ mode, and therefore, as we move from the basic cut to the $\chi^2$ cut, the $AAW$ mode dominates over the $AAAW$ mode \footnote{While the $AAAW$ mode can give rise to the $6b + \ell + {\rlap{,/}{E}_T}$ final state, tagging $6b$-jets, especially for background events, is extremely challenging. Hence, we refrain from pursuing studies in this direction.}. Comparing the signal and the background, we can clearly see that the $\chi^2$ cut kills the background significantly. Thus from our analysis with six BPs selected over a wide range of masses, we demonstrate that the signal significance greater than $3\sigma$ can be achieved. 

For our analysis we considered only two scenarios for the pseudoscalar mass: 50 GeV and 70 GeV. The former scenario is chosen to explore the situation when $h\to AA$ is possible, while the second scenario is chosen where $h\to AA$ is not allowed. Additionally, one could also consider scenarios with a slightly higher mass for $A$, such as $m_A = 110$ GeV and the remaining BSM Higgs bosons heavier than $A$. In that case, BR($A \to b\bar{b}$) is approximately $70\%$, and we expect good results for such scenarios as well. However, it should be noted that the EW production cross section of the $AAW$ mode will decrease as the cumulative mass of $H^\pm$ and $A$ increases. Consequently, the significance will be relatively low for higher masses of $H^\pm$. For $m_A > m_{{h, H, H^\pm}}$, the bosonic decay modes of $A$, such as $A \to hZ^{(*)}/HZ^{(*)}/H^\pm W^{(*)}$, will become accessible and may dominate over the $A\to b\bar{b}$ mode in the fermiophobic limit (large $\tb$). Thus, in the context of the $4b + \ell + {\rlap{,/}{E}_T}$ final state in type I 2HDM with standard mass hierarchy, choosing the pseudoscalar as the lightest of all the Higgs bosons would be appropriate.

 
Before we conclude, we would like to point out how the $\chi^2$ discriminator can be applied to real data at the LHC as a selection criteria in the context of BSM Higgs boson searches. The experimentalists should examine the normalized $\chi^2$ distributions using real data. When the correct hypothesis (masses of $A$ and $H^\pm$) is used as input to the $\chi^2$ discriminator, the resulting $\chi^2$ distribution will exhibit a declining pattern with respect to $\chi^2$, similar to the signals shown in Fig.[\ref{fig: chi_square}]. Concurrently, experimentalists should vary (or make a proper scan) the masses of $A$ and $H^\pm$ to determine the values that yield the steepest decline in the $\chi^2$ distribution. These identified masses of $A$ and $H^\pm$ represent the correct masses of $A$ and $H^\pm$ that exist within the signal. Not only that, for the correctly identified masses which shows the steepest decline in the $\chi^2$ distribution, the $\chi^2$ cut would give very high discovery significance.

\section{Conclusions}
\label{sec:conclusion}

The EW multi-Higgs production in various BSM frameworks can be dominant compared to the QCD induced processes due to the non-standard couplings of the additional Higgs bosons. In type I 2HDM, for a large region of parameter space, all the additional Higgses exhibit fermiophobic behaviour and hence, the EW processes are dominant. Not only that, the EW processes can also be $q\bar{q}'$ induced, which results into charged final states like $p p \to H^\pm A \to AAW$ and therefore $4b + \ell + {\rlap{\,/}{E}_T}$ final state for light pseudoscalar at the LHC which cannot be achieved through QCD processes. Thus, alongside the QCD induced processes, systematic analysis based on the EW processes should be done both phenomenologically and experimentally. 

In this work we studied the $4b+ \ell + {\rlap{\,/}{E}_T}$ final state at 13 TeV LHC with a luminosity of $3000$ fb$^{-1}$. However, the signal is obscured by large $t\bar{t}+$jets background. Hence strong selection cuts are to be imposed to kill the background without much affecting the signal. The dominant contribution to the $4b+ \ell + {\rlap{\,/}{E}_T}$ final state at the generator level comes through the $AAW$ mode, we use this signal topology and the  BSM Higgs mass informations like the masses of $A$ and $H^\pm$ (signal hypothesis) to construct the $\chi^2$ variable. We briefly summarize the $\chi^2$ method in the following steps:

\begin{enumerate}
\item First we imposed a $b$-jet pairing algorithm to make two $b$-jet pairs out of the four leading $b$-jets.

\item Assuming that the MET corresponds to only one neutrino from the leptonic decay of the $W$ boson in the $AAW$ mode, we estimated the $p_{z_\nu}$ component of neutrino momentum.

\item We did the truth reconstruction of the BSM Higgs bosons ($A$ and $H^\pm$) involved in the $AAW$ mode to obtain the mass resolutions. These mass resolutions are used as inputs to the $\chi^2$.

\item Finally, we make all possible combinations of $\chi^2$s and pick the one which is minimum as the $\chi^2$ of the event.   
\end{enumerate}

 The $\chi^2$ serves as a powerful tool to discriminate the signal from the background. We showed that the $\chi^2$ distribution for the $AAW$ signal falls very sharply with $\chi^2$ whereas the background distribution is very broad. Our formalizm for $\chi^2$ is appropriate only for on-shell $W$ boson and therefore scenarios restricted to on-shell $H^\pm \to AW$ processes and also for $W$ boson decaying to the leptonic state with one neutrino. Since the contribution of more than one neutrino and electron/muon via taonic decay of $W$ boson is negligibly small, we can safely apply the $\chi^2$ method even if the signal is generated with $W \to \ell \nu$ and $\ell$ includes $\tau$ lepton.  We impose  $\chi^2 < 1$ and other selection cuts like asymmetry cut and di-jet $\eta$ separation cut to effectively reduce the background. To study the discovery prospects of $4b+ \ell + {\rlap{\,/}{E}_T}$ final state at the LHC, we considered the subdominant EW processes along with the primary $AAW$ mode and obtained the discovery significance greater than $3\sigma$ over a wide range of parameter space. 
  \\


\acknowledgments
 
 The authors thank Ravindra K. Verma for some useful discussions. The authors would also like to thank Stefano Moretti and Jeonghyeon Song for careful reading and useful comments. The work is supported by the National Research Foundation of Korea, Grant No. NRF-
2022R1A2C1007583.




\bibliography{ref}
\end{document}